\documentclass[conference,compsoc]{IEEEtran}

\usepackage[nocompress]{cite}
\usepackage{url}
\usepackage{array}
\usepackage[caption=false,font=footnotesize,labelfont=sf,textfont=sf]{subfig}
\usepackage{fixltx2e}
\usepackage[pdftex]{graphicx}
\usepackage{multirow}
\usepackage{lipsum}
\graphicspath{{img/}}
\DeclareGraphicsExtensions{.pdf,.jpg,.png}

\begin{document}

\title{Survey of robust and resilient social media tools on Android}

\author{
  \IEEEauthorblockN{Paul Brussee}
  \IEEEauthorblockA{Delft University of Technology\\Email: P.W.G.Brussee@student.tudelft.nl}
  \and
  \IEEEauthorblockN{Johan Pouwelse}
  \IEEEauthorblockA{Delft University of Technology\\Email: J.A.Pouwelse@tudelft.nl}
}

\maketitle

\begin{abstract}
We present an overview of robust and resilient social media tools to overcome natural disasters, censorship and Internet kill switches.
These social media tools use Android devices to communicate during disasters and aim to overcome attacks on freedom of expression.
There is an abundance of projects that aim to provide resilient communication, enhance privacy, and provide anonymity.
We focus specifically on the limited set of mature tools with a healthy development community and Internet-deployment.
\end{abstract}


\section{Introduction}

Social media has become a widespread phenomenon of local, national and global communication between people.
This free human communication via social media is under threat of blackouts and far-reaching censorship.
Recent natural disasters have wiped out vital communication infrastructure for a prolonged period of time \cite{renesys2005katrina} or caused a system overload due to the way the network is structured with central bottlenecks that are also a single point of failure \cite{denis2010haiti}.
Communication networks can also be unreliable due to a suddenly unstable power grid \cite{cnn2003poweroutage} or permanent power shortages \cite{nyt2015afrika}.

Censorship can be effectuated in such a way that it resembles partial network failures \cite{hrw2006china} or a total blackout \cite{watts2014havana} with an Internet kill switch \cite{passary2015socialmedia}.
A constant battle is fought over tools that counter a specific censorship mechanism and vice versa \cite{halderman2013iran}.
State actors are involved in what seems to be a direct battle between them and the open source community providing activists with the tools they need \cite{nesterov2015tor}.

Mobile phones can be considered autonomous platforms thanks to wireless ad hoc connections and on board rechargeable batteries \cite{quartey2015phone}.
The smart phone app phenomenon has been used as an angle to counter disruptions in communications due to natural disasters or human made interference \cite{brussee2015autonomous}.
This recently prompted even more aggressive censorship measures behind the Great Firewall of China \cite{nyt2015china}.

\subsection*{Problem Description}

This survey lists field proven solutions that provide a robust and resilient social media experience on Android, one of the most popular mobile platforms, considering privacy and anonymity.
The threat model assumed is the following: \cite{pouwelse2012censorshipfree}
\begin{itemize}
\item{The adversary can observe, block, delay, replay, and modify traffic on all underlying transport. Thus, the physical layer is insecure.}
\item{The adversary has a limited ability to compromise smart phones or other participating devices. If a device is compromised, the adversary can access any information held in the device\'s volatile memory or persistent storage.}
\item{The adversary cannot break standard cryptographic primitives, such as block ciphers and message-authentication codes.}
\end{itemize}

\section{The Twimight Ad Hoc Network}

\begin{figure}[b!]
\centering
\includegraphics[width=3in]{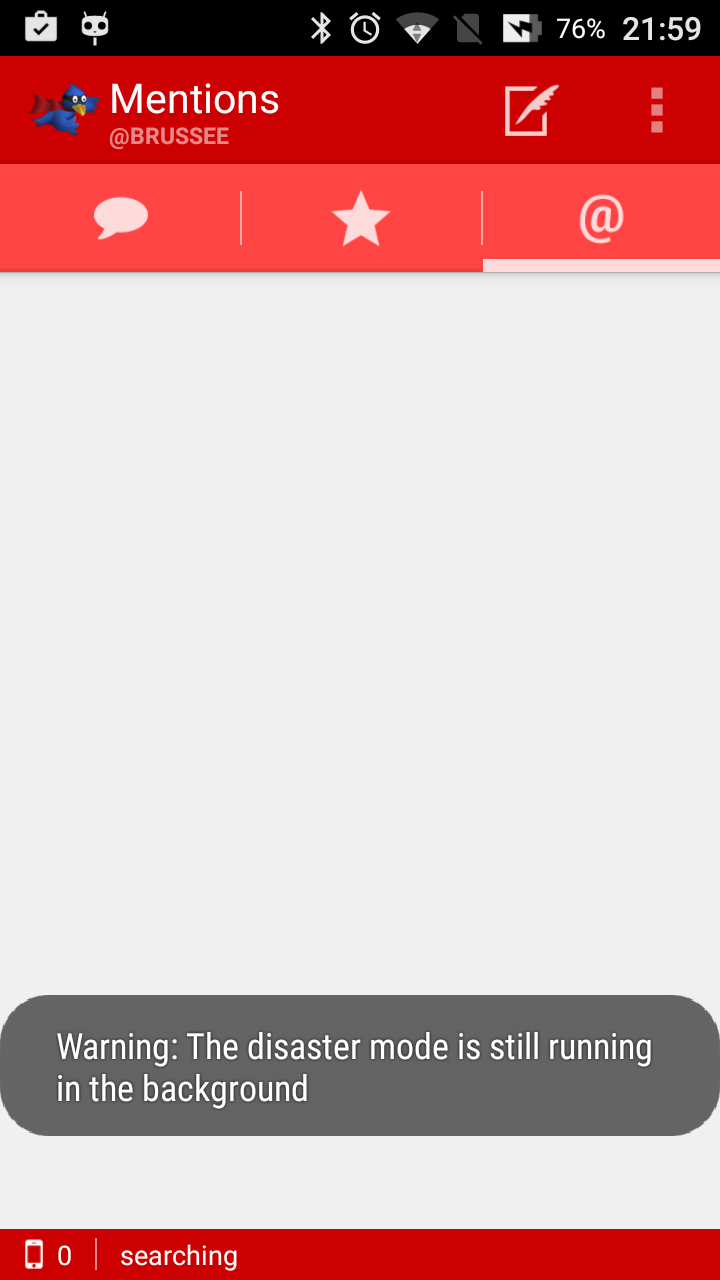}
\caption{The Twimight home screen.}
\label{fig_twimight}
\end{figure}

Twimight uses opportunistic ad hoc communication and the store and forward principle to distribute tweets amoung users with the same Twimight app in 'disaster mode'.
\cite{Hossmann:2011:TDM:2079360.2079367}
In normal mode it behaves just like a regular Twitter app authorized by the user with OAuth.
In disaster mode Twimight tries to spread tweets epidemically and publicly to every peer within reach, flooding the network.
Switching to this disaster mode enables Bluetooth to periodically scan for other devices, also running the Twimight app in disaster mode, while continuously listening to answer connection requests from other devices.
Upon finding a peer the Bluetooth connection is automatically setup and 'disaster tweets' are exchanged.

The Twimight Disaster Server (TDS) is a central component in the system that signs a certificate for each Twimight user.
This certificate is used to sign disaster tweets for integrity checks and author authentication, as well as to encrypt private direct messages between clients.
The lifetime of each certificate is limited to only 14 days to keep the certificate revocation list embedded in the Twimight app short. \cite{Hossmann:2011:TDM:2079360.2079367}

To prevent expiration of a certificate just before a disaster happens the clients update their keys and request a new certificate from the TDS after 7 days.
If the duration of a disaster is more than 7 days the authors of Twimight expect an alternative infrastructure to be available at that time to access the Internet. \cite{Hossmann:2011:TDM:2079360.2079367}

New users can not join the network during a disaster because it is impossible to use the disaster mode without a TDS signed certificate.
The TDS is the only allowed certificate authority by the Twimight app and all invalid tweets are discarded.
The TDS root certificate is embedded in the Twimight app for validation of every certificate that is used to sign a tweet.
To update the root certificate requires the Twimight app to be updated. \cite{Hossmann:2011:TDM:2079360.2079367}

To control spam the maximum of tweets exchanged is 500 of which at most 250 are from the peer itself and the rest relayed from others.
It is expected of the user to moderate the amount of their own tweets and to retweet only relevant information.
There is no integrated resource management strategy in case this buffer gets full. \cite{Hossmann:2011:TDM:2079360.2079367}

If Internet connectivity is regained only the user's own disaster tweets are uploaded to the Twitter server \cite{Hossmann:2011:TDM:2079360.2079367}.
In order for a disaster tweet of a user that may still be in harm's way to reach the Internet requires a retweet from a user with Internet connection.
Twimight is not able to check if the disaster tweet from another user is already published and cannot decide to publish this tweet on behalf of this user, who is apparently still without Internet connection, and who possibly remains in harm's way.

The opportunistic network approach requires continuous scanning for connection opportunities.
A shorter interval of 1 minute versus 2 minutes using Bluetooth severely degrades battery life.
A larger interval may result in missed connection opportunities and resulting consequences in the context of a disaster.

\section{The Serval Mesh Network}

\begin{figure}[b!]
\centering
\includegraphics[width=3in]{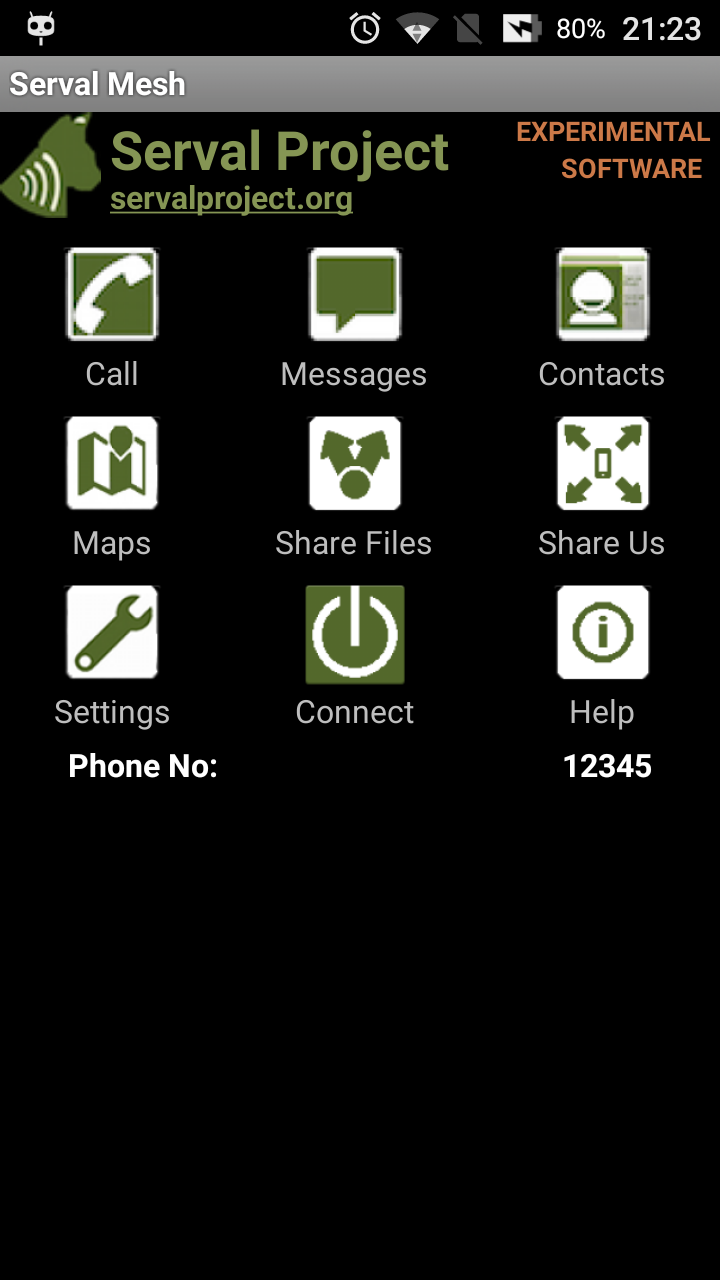}
\caption{The Serval home screen.}
\label{fig_serval}
\end{figure}

The Serval Project created mesh network software to provide communication services without the need for fixed infrastructure.
Smart phones equipped with WiFi and the Serval Mesh app for Android allow voice calls to be made over the mesh network as well as text messaging and file sharing.
Within any mobile network the clients can physically change location and therefore change location in the network.
Particularly in mesh networks there is no central directory to register the new location of a client, and the network itself may change in addition to that, making routing even more complex.
To connect to a specific user requires knowing their phone number and asking all neighboring phone in the mesh network to find the address of the phone on the network.
The Mesh Datagram Protocol (MDP) in the network layer propagates these requests over the network for a limited time and returns the addresses of any phone that claims to own that phone number.
It is impossible to verify these claims without prior knowledge of the public key of the callee so it is left up to the user to choose the right result, without knowing if the intended receiver was actually reached by the request at all.
Each user can claim their own phone number on the mesh without any checks and generate their own public-private key pair with elliptic curve cryptography (ECC).
The public key is used as unique subscriber identifier or Serval Identity (SID) and both keys are stored on the device and hidden using steganography for plausible deniability  \cite{bettison2014servaldna}.
Only after the first contact a certificate can be issued by the caller to the callee to affirm their identity and build a trust network.
Before knowing the public key of the intended receiver anyone in between the connection can impersonate and eavesdrop on the conversation.
A man-in-the-middle attack is generally very difficult to defeat in a mesh network because often no trusted third party can be contacted to verify the validity of a certificate.

Another pitfall of mesh networks is the usual high end-to-end packet loss.
To reduce but not guarantee no end-to-end packet loss MDP uses per-hop retransmission which restores otherwise unusable routes and provides better packet delivery rates than conventional UDP over a multi-hop wireless network.
MDP also allows for packets to be signed and encrypted so upper layers don't have to.
Directly above the network layer is the voice call session and audio carrier protocol called  the Voice over Mesh Protocol (VoMP).
VoMP is more suitable for mesh networks as it does not assume stable virtual circuits over static routes, like SIP and RTP do.
Since MDP does not offer reliable messaging VoMP is designed to handle packet loss, mid-call re-routing and re-connection where SIP would fail. \cite{serval2014dna}

The resilient file distribution system Rhizome is used for file sharing.
Rhizome uses the store and forward principle and is therefore suitable for mesh networks, but only guarantees the integrity of the file and not eventual arrival.
The best effort transport of files is prioritized according to network topology, bandwidth and device storage and battery availability, but otherwise it preserves net neutrality.
Anonymous file sharing is possible by not storing any information on the senders so they can not be traced. \cite{bettison2014servaldna}
The messaging service MeshMS is build upon Rhizome by sharing two journal files per interlocutor, one for incoming and one for outgoing messages.

To overcome the scalability limitations of mesh networks with link tracking, like the Serval Mesh, route aggregation is necessary beyond the point of all bandwidth being used for link tracking.
Network prefixes based on physical or topological location are proposed as well as making the network interoperable with infrastructure like the  Commotion wireless network \cite{commotion2012mesh}.
Any device that does not know its own location can clone a nearby network prefix and provide a coarse location service.
Anonymity however suffers from this approach.

\section{The Psiphon Proxy \& VPN Service}

\begin{figure}[b!]
\centering
\includegraphics[width=3in]{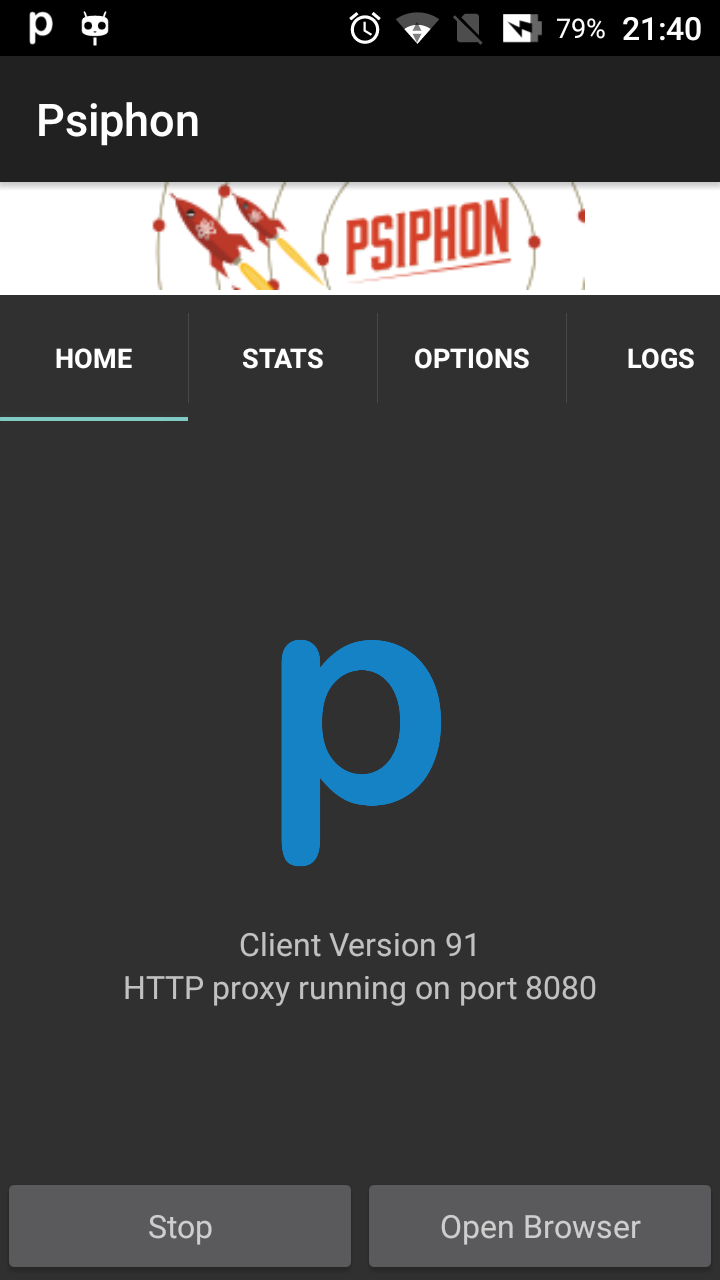}
\caption{The Psiphon home screen.}
\label{fig_psiphon}
\end{figure}

Psiphon is a proxy client and server tool originally made in 2006 by the Citizen Lab in the OpenNet Initiative and now developed by Psiphon Inc.
It is used in more than 40 countries where Internet is censored in some way \cite{hoag2012syria}.
The tool allows users to circumvent censorship by connecting to a proxy server outside of the censor's network and create an encrypted tunnel through which the rest of the Internet can be accessed.
Psiphon uses the socket secure (SOCKS) and  Secure Shell (SSH) protocol and virtual private network (VPN) to implement the secure tunnel as well as HTTP proxy technology.
These protocols provide an encrypted end-to-end tunnel that prevents the censor from knowing the contents of the communication providing some privacy.
An obfuscation layer can be added on top of SSH in an attempt to circumvent deep packet inspection (DPI) that can recognize the type of traffic and block it regardless of its content and destination.
The tunnel goes straight from the client to the proxy server and no effort is made to hide the client's IP address, so Psiphon does not provide any anonymity at all.
This important dissent feature is deemed out of scope \cite{isec2014psiphon}.

Keeping Psiphon operational requires new proxy servers to be available at the same rate or quicker than used proxy servers are blocked.
If the censor gets ahead in this cat-and-mouse game the Psiphon client will no longer able to connect to any proxy server and is forced to find another source for addresses of non-blocked proxy servers in the future.
Domain fronting is used to hide the true destination of an HTTPS request by putting a fake front domain in the DNS query and in the TLS Server Name Indication (SNI) extension, and the real one in the HTTP Host header \cite{fifield2015domain}.
Using content delivery networks (CDN) to host the domain of the proxy server is an attempt to make the cost of blocking Psiphon together with the entire CDN too high for the censor.
Accessing the proxy by IP directly, like domainless fronting, leaves the option of blocking the IP address, and with it all domains and services behind it.
The amount of available addresses is limited and will eventually deplete the remaining stack of IPv4 addresses.
Psiphon is IPv6 compatible but the censor can block these addresses just as easily when the censoring tools are IPv6 compatible as well.

Psiphon is vulnerable to an insider attack because the censor can use the tool itself to discover all non-blocked proxy servers and subsequently block them.
Various strategies based on software build ID, date, and time of day, are used for publishing a limited amount of servers to a single client to mitigate this risk of an insider attack.
The idea of tracking which clients know about which servers to find traitors was rejected because that requires tracking with an unique client ID \cite{psiphon2011design}.

Certificates are used to verify the authenticity of the clients and servers.
Proxy servers are authenticated by the client automatically with embedded certificates in the software to make it hard for the censor to perform a man-in-the-middle attack.
Because authenticating the client is a manual process that requires some technical knowledge it undermines the security of the embedded server certificates.
A malicious repackaged version of Psiphon surfaced in 2014, that was likely part of a targeted attack by actors in Syria, installed a remote access tool (RAT) and key logger \cite{scott2014malicious}.
Psiphon Inc. is a commercial company and it collects and sells the metadata of its users, for example which sites are visited from which countries and how often \cite{psiphon2015website}.
This logging may concern anyone that uses Psiphon to circumvent censorship if the act of doing so is illegal.
Also the company allows sponsors to automatically open a website in the user's browser after a tunneled connection is established which makes a individual user traceable.

\section{Orbot Onion Routing Service}

\begin{figure}[b!]
\centering
\includegraphics[width=3in]{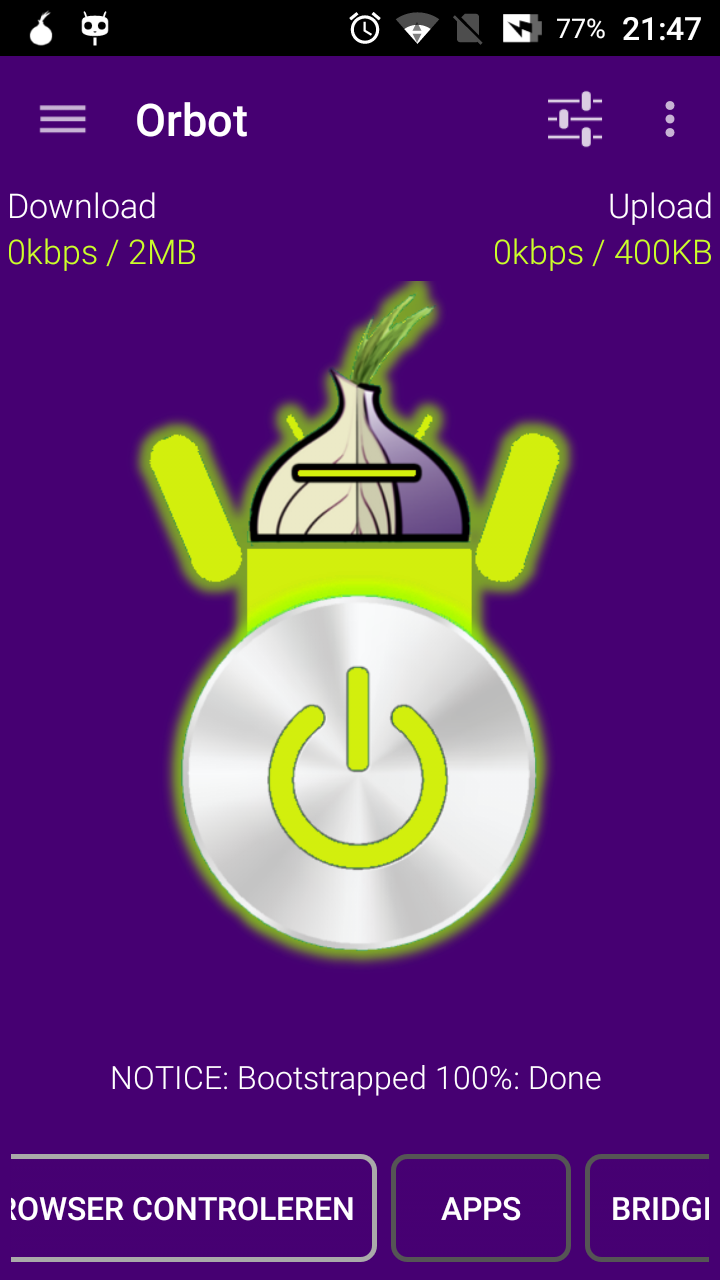}
\caption{The Orbot home screen.}
\label{fig_orbot}
\end{figure}

The second generation onion router (Tor) is a circuit-based privacy-enhancing communication service \cite{dingledine2004tor}.
The distributed overlay network Tor creates is designed to anonymize TCP-based applications like web browsing.
Orweb is such a web browser and uses the Orbot app as the Tor client software.
Tor clients build temporary circuits from available relay routers that are found in a Tor directory server.
A typical circuit has a fixed length of three different nodes: 1. entry node 2. relay node 3. exit node.
The initiator chooses which nodes to use and creates a tunnel through each, ending up with three layers of encryption.
The nodes only know their direct neighbor in the circuit, so operate anonymously beyond their direct neighbor, and are not supposed to log this information to enhance privacy.
Anonymity outside of the Tor network is the result of the exit node connecting outside of the Tor network on behalf of the initiator.
From the exit node onwards the connection is no longer secured by Tor.
To prevent abuse not every exit node allows all traffic and all destinations and some only connect to a private network.
To achieve low latency the initiator makes use of the advertised available bandwidth for a relay router to do load balancing. \cite{Bauer:2007:LRA:1314333.1314336}
Perfect forward secrecy is achieved by negotiating session keys for each successive hop in the circuit.

The directory server is a central component in the Tor network and is the authority of the list of relays that make up the Tor network.
Only 10 directory servers are currently in existence and all the addresses are hard coded into the Tor Browser Bundle.
The directory server must approve every new onion router and must be trusted by all its users.
Bridge relays however are not listed in the main Tor directory making it difficult for a censor to block all known Tor relays.
Limited amount of bridge addresses are provided upon manual user request by the bridge authority via email \cite{tor2015bridges}.
Obfuscation protocols are used to connect to the first Tor relay and constantly improved to stay ahead of the deep packet inspection of censoring ISPs.

Inside its network Tor offers location-hidden TCP-services, like a web server.
The service provider can stay anonymous by creating circuits to multiple introduction points that can be contacted by potential users of the service.
The user collects the onion identifier of the hidden service and create its own circuit to the introduction point as well as a separate circuit to another relay router that acts as a rendezvous point.
The hidden service learns of the user's request from its introduction point and connects via another circuit to the rendezvous point where the two circuits are mated.
Various attacks \cite{4725864} \cite{6682740} sometimes involve just one malicious node to compromise anonymity of a hidden service within minutes or hours \cite{1624004}. 
An active attacker can always try to analyze the timing and volume of traffic going in and out of the relay to de-anonymize its target due to the low-latency design of the Tor network. \cite{Tang:2010:IAT:1866307.1866345}
To counter this attack angle a relay that is deemed reliable in terms of uptime and capable in terms of bandwidth availability receives the Guard flag.
A small subset of guard nodes are selected by a Tor client for a longer period of time to choose entry nodes from.
If an adversary is not in this subset of entry guards it prevents the adversary from eventually becoming the first relay in the circuit after frustrating the circuit building process of the Tor client.
This gives the Tor user a chance to not be observed from the first relay, that is the only relay that knows the true origin.
Although guard nodes are supposed to guard against path disruption attacks, they are themselves vulnerable to attack \cite{Bauer:2007:LRA:1314333.1314336}.

\section{The Tribler Family}

Tribler is a fully decentralized peer-to-peer social sharing app that aims to provide a Youtube-like video consuming experience with an integrated multimedia player.
On the desktop Tribler is a single application which runs on all three major platforms, but on Android the core components and interface are not yet well integrated.
On the mobile platform the Tribler family of apps offer additional privacy enhancing features, viral spreading and mutation from source code.

The two authors of this survey are doing their thesis work on Tribler and founded Tribler respectively.

\subsection{Tribler Play}

\begin{figure}[b!]
\centering
\includegraphics[width=3in]{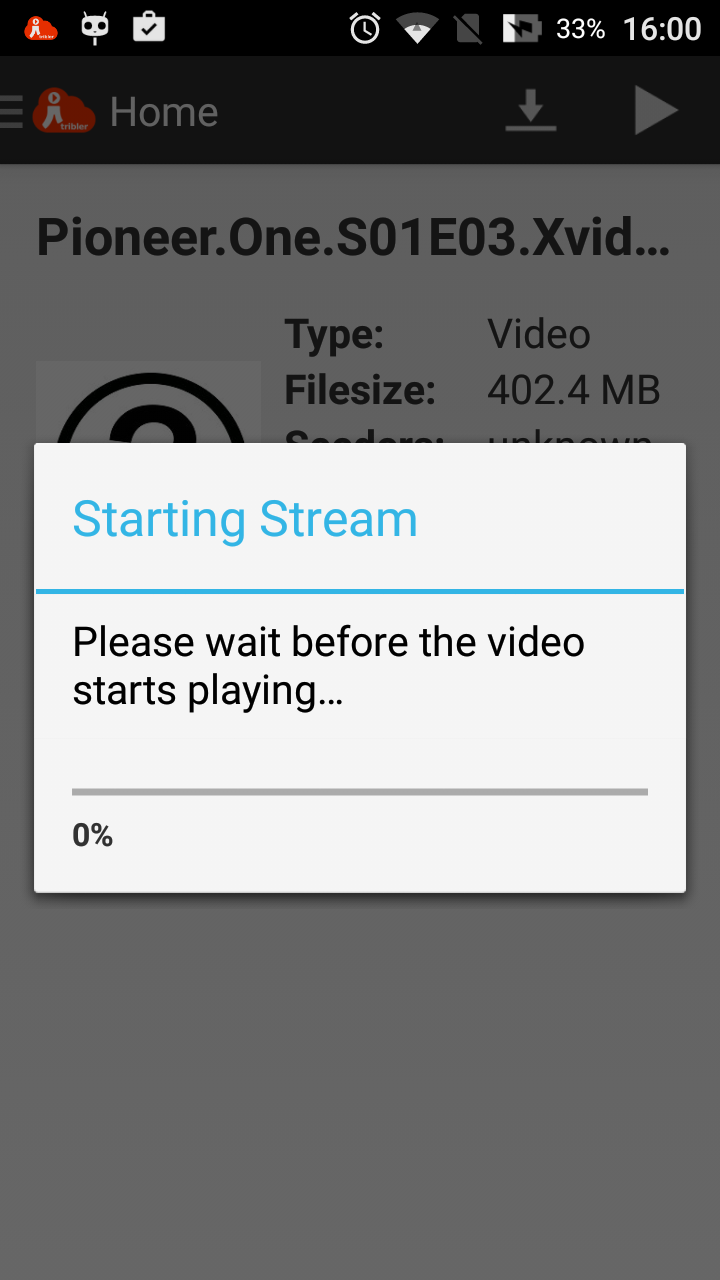}
\caption{The Tribler Play home screen while starting a stream.}
\label{fig_triblerplay}
\end{figure}

The Tribler Play app consists of a native Android Java GUI, VLC for Android player and Python for Android to run Tribler underneath. \cite{tribler2014play}

Tribler is capable of distributed streaming of real-time HD video and efficiently adapt to network conditions \cite{Pouwelse:2004b, pcs2004}.
It is based upon and compatible with the BitTorrent protocol, but without the need for any central component in the system, like trackers or websites.
Search functionality is integrated into the app by the elastic database Dispersy.
Content is announced on channels, just like Youtube, and each user can create one.
Tribler uses three specific Dispersy communities to distribute torrent files and thereby replace the typical BitTorrent websites: AllChannelCommunity, SearchCommunity and MetadataCommunity.
Deep knowledge and performance analysis of the BitTorrent protocol generated insights for improvements that were applied in Tribler \cite{Pouwelse:2004, iptps2005, aiosup05bittorrent, WeticeBitCrunch}.
For instance, timely and accurate knowledge of the geographical location of peers in the network can increase the transfer speed significantly \cite{aiosup05geo}.
In order to propagate and store metadata a gossip type protocol is used \cite{Wang:SIGIR2007, votecast}.
A study of the top-10 Youtube content creators indicate that essential features for effective platforms for peer production are: separation of good and bad contributions, regulation of computer resources, group communication, and community building \cite{telcom-policy-2008}.

Building upon social relationships Tribler gives content recommendations to the user automatically based on the preferences of friends and people with similar interests and improves these suggestions even without interaction \cite{beyond2005, ASCI-2008-buddycast}.
Self-organizing collaborative filtering can improve these recommendations even without interaction \cite{Wang:WiFiWalkman, Wang:SelfOrganizing, Wang:Distributed_ASCI, ClementsIR2007, Wang:2008a}

Free-riding can not be prevented if more than half the peers are behind a firewall, since they are unable to respond to incoming connections \cite{JanDavidFFFPaperIEEE2008}.
Therefore to be able to provide a reasonable quality of service for fair users, free-riders in the system are penalized as long as upload bandwidth in the swarm is scarce \cite{mol2008a}.
A distinction is made between lazy free-riders and die-hard free-riders, the latter of which cheats the protocol \cite{bartercast}.

Tribler correlates multiple overlay networks to characterize the behavior of real users \cite{mprobe05gp2pc}.
For example, communities of users with similar interests can be exploited to automatically build a robust semantic and social overlay to increase usability and download performance significantly by collaborative downloading \cite{TriblerOverviewJournal, PawelOffloading}.
The social network can also increase trust and give an incentive to exchange content which increases content availability \cite{Wang:2008a}.
Since cooperation is hard to enforce Tribler uses psychology theory for inducing it by showing the similarity between users, making behavior
publicly visible, making it possible for users to express approval or disapproval
about each other, allowing different return-on-investment times for different relationship
types, showing how much others contribute in respect to their total resources, exploiting the
need to belong, and make groups small, powerful and exclusive, allowing users to stand
out of the crowd, displaying information on friendship and the degrees of separation, letting 
users specialize in one of the four aspects of cooperation, keeping track of the given
help/received help ratio, making information about a group and its members visible, and
finally, making users feel intrinsically motivated to cooperate \cite{FokkerEuroITV2007, fokker2007b}.
A user is more induced to reciprocity over time if the action is directed at one person rather than whole community as a whole \cite{fokker2008a}.
Effective reciprocal relationships can most likely be found within first or second degrees of separation, whereas relationships of higher degrees are not as flexible over time, thus benefiting most from immediate reciprocity. \cite{DelayedReciprocity2008}

\subsection{Android Tor Tribler Tunneling (AT3)}

Android Tor Tribler Tunneling (AT3) is a research prototype that proves the feasibility of using anonymous tunnels on an Android device through which Tribler can offer its services to the user as long as the Android device supports hardware accelerated cryptography or encryption is disabled.
AT3 is the sibling of Tribler Play, both use Python for Android to run the Tribler core and an Android port of the BitTorrent library (libtorrent), and the apps together offer the same functionality as the Tribler desktop application. \cite{triber2014at3}

The Tor-like anonymity network is very similar to plain Tor, however it is a separate network and allows multiple circuits at the same time to be aggregated to increase download speed.
The app anonymously downloads a torrent of 50 MB over 1-3 hops and 1-3 circuits while measuring CPU usage and download speed, averaging 450-650 KB/s \cite{triber2014at3}.

The regular BitTorrent protocol on top of plain Tor leaks information and becomes more vulnerable to traffic monitoring and communications' hijacking \cite{DBLP:journals/corr/abs-1004-1461}.
Tribler however eliminated the tracker that would normally reveal the true IP addresses of its users, thus Tribler now combines the best of both worlds.

\begin{figure}[h!]
\centering
\includegraphics[width=3in]{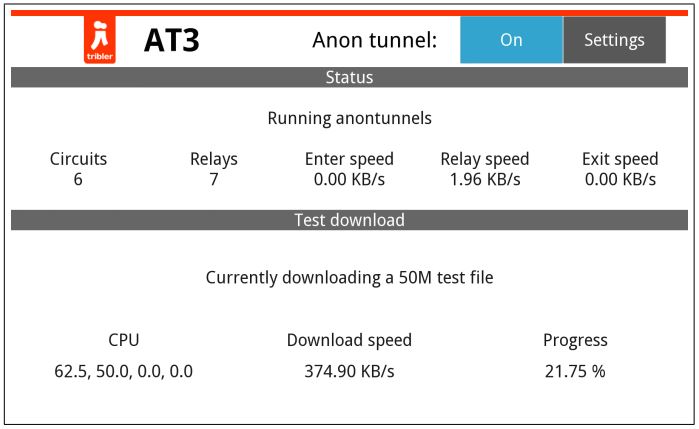}
\caption{The Android Tor Tribler Tunneling home screen.}
\label{fig_at3}
\end{figure}

\subsection{Shadow Internet}

\begin{figure}[b!]
\centering
{
\setlength{\fboxsep}{0pt}
\setlength{\fboxrule}{1pt}
\fbox{\includegraphics[width=3in]{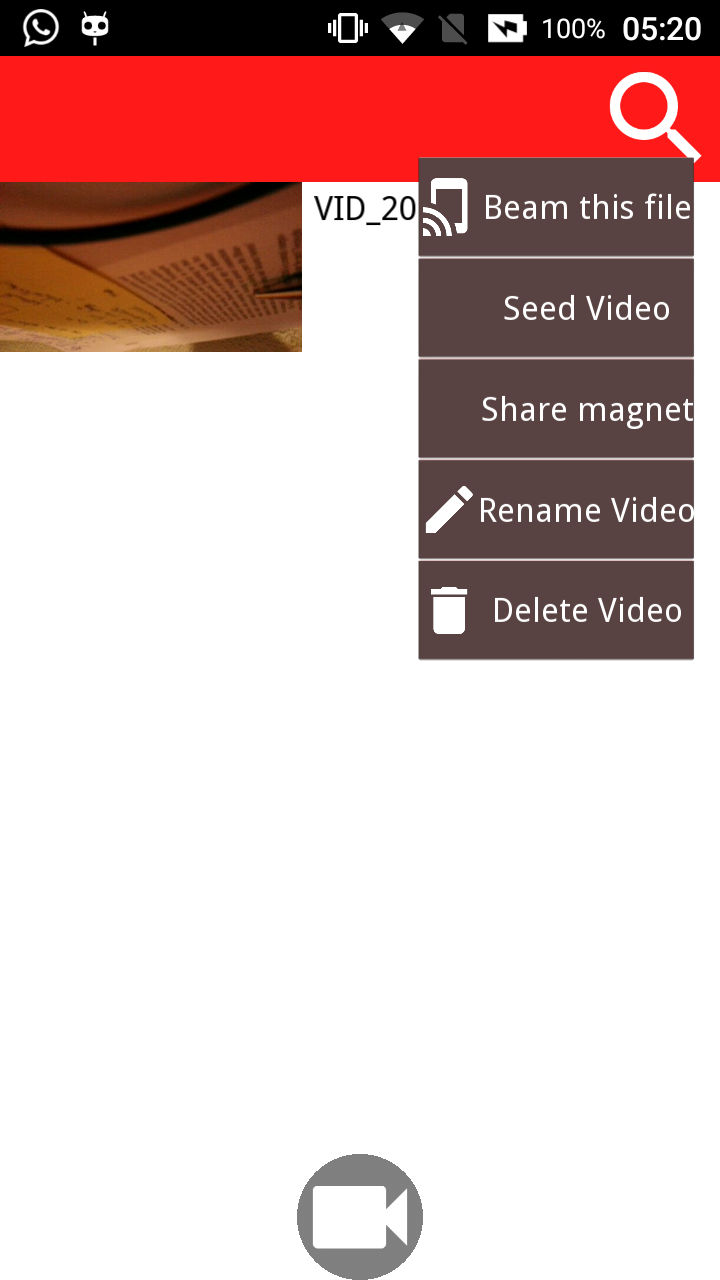}}
}
\caption{The Shadow Internet home screen with video context menu.}
\label{fig_shadowinternet}
\end{figure}

The Shadow Internet app currently integrates a video recording feature, NFC+Bluetooth wireless app \& file transfer feature with the Tribler search and download features.
The entire app, including the interface, are written in Python, supported by Python for Android. \cite{tribler2015shadowinternet}

Android smart phones that are equipped with a Near Field Communication (NFC) chip can use this to automatically setup a Bluetooth large file transfer with the Android Beam feature.
The Shadow Internet app can transfer the videos made with the phone as well as the app itself.
Leveraging the ease of use and power efficiency of NFC+Bluetooth transfers the smart phone is turned into a node of an ad hoc network wherever Internet is not available.
When Internet is available the app would use the Tribler network to seed the video, but is currently not yet implemented.

In the use case of eye witness of government misbehavior the app would allow the user to record and share the evidence anonymously directly from the Shadow Internet app.
Spreading the data as fast as possible and possibly getting it out of the government controlled region is key for the evidence to come to light.

Another use case is the festival fanatic who records a video of the experience and wants to instantly share it with friends in an extremely busy area or an area without Internet access.
In both instances the wireless NFC+Bluetooth transfer allows the user to share the content to anyone without the app even.
If a WiFi connection is available the Tribler protocol would publish and distribute the video effectively via the users' channel.

\begin{table*}
\renewcommand{\arraystretch}{1.4}
\centering
\begin{tabular}{l|*{9}{r}}
Name	&	Age	&	Last	&	Commits	&	LOC	&	\# Contributers		&	\# Open	&	\# Open	&	Min. Android	&	\# Installations	\\
	&		&	Release	&		&		&		&	Pull Requests	&	Issues	&	Version	&	\\
\hline
Psiphon	&	9 y	&	nov `15	&	6 K	&	363 K	&	29	&	4	&	174	&	2.2+	&	10.000.000 - 50.000.000	\\
Tor		&	13 y	&	oct `14	&	46 K	&	435 K	&	287	&	-	&	2425	&	2.2+	&	5.000.000 - 10.000.000	\\
Tribler	&	10 y	&	jan `15	&	15 K	&	166 K	&	72	&	8	&	186	&	4.0+	&	1.000.000 - 5.000.000		\\
Serval	&	11 y	&	jul `15	&	12 K	&	420 K	&	120	&	8	&	17	&	2.2+	&	100.000 - 500.000		\\
Twimight	&	4 y	&	sep `14	&	0.5 K	&	24 K		&	6	&	-	&	9	&	4.0+	&	5.000 - 10.000			\\
\end{tabular}
\vspace{5pt}
\caption{Comparing projects \cite{redecentralize2015alternativeinternet}.}
\label{tab_compareproj}
\end{table*}

\subsection{DroidStealth}

\begin{figure}[b!]
\centering
\includegraphics[width=3in]{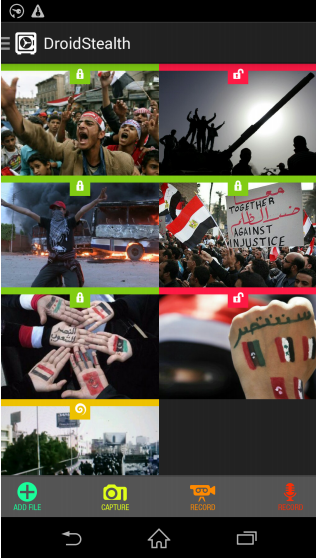}
\caption{The DroidStealth home screen with content.}
\label{fig_droidstealth}
\end{figure}

The DroidStealth app has a morphing capabilities, cryptography containers and can hide traces of usage of itself on the phone.
The app acts like a vault in which content can be securely saved thanks to encryption.
Just like the Shadow Internet app \cite{tribler2015shadowinternet} it also features a video camera recorder and on top of that a voice and photo capture feature. \cite{hokke2015droidstealth, tudelft2014droidstealth}
The user interface is very intuitive thanks to an icon and color indication of which content is currently (not) encrypted, and simply tapping to toggle the status.
The most impressive features are the hiding capabilities that do not require special permissions on the device: morphing the icon and name to something inconspicuous, the hidden launchers, and automatic removal of traces of usage.

The app can be secretly launched by typing a user configured pin-code into phone dialer or with an invisible widget.
The fake phone number won't be called and won't show up in the recent called list.
After switching to another app or the desktop DroidStealth is no longer present in the recent app carousel.

The morphing is achieved through unpacking the Android Application Package (.apk) and modifying the binary xml files and icon.png, then repackaging and signing with an embedded key.
The morphed app can then be shared with others via NFC+Bluetooth just like the Shadow Internet app \cite{tribler2015shadowinternet}.

Any adversary that performs a casual check on the phone will not notice a calculator or another similarly innocuous app.
All suspect traces of usage are removed automatically by the app and it can superficially look like any other app with extreme ease.

\subsection{SelfCompileApp}

\begin{figure}[b!]
\centering
\includegraphics[width=3in]{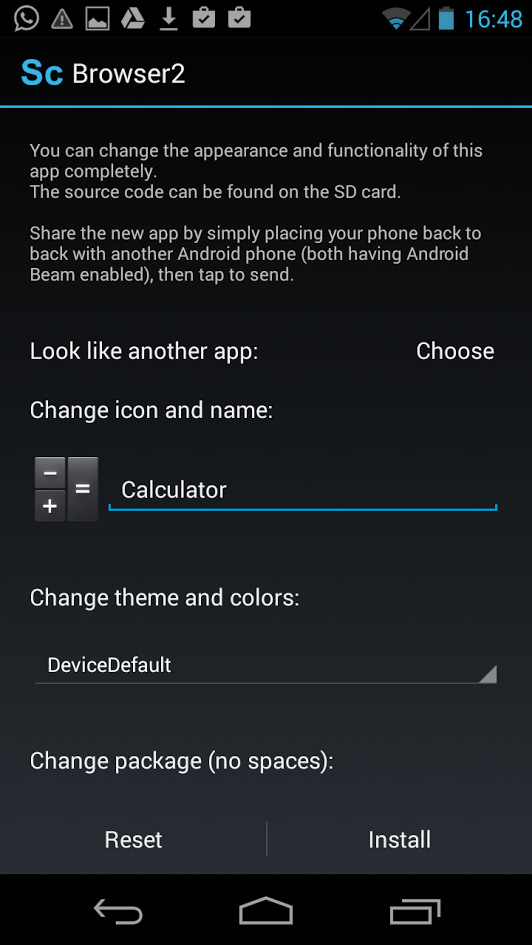}
\caption{The SelfCompileApp home screen.}
\label{fig_selfcompileapp}
\end{figure}

The SelfCompileApp is the first autonomous app that is capable of self-compilation, mutation and viral spreading \cite{brussee2015autonomous}.
It does not require a host computer to alter its functionality or change its appearance, and no special permissions on the device.
It also does not need to be distributed through an app store or website because it has the same NFC+Bluetooth transfer feature as DroidStealth \cite{hokke2015droidstealth}.
It has all the required Android SDK build tools embedded and a Java key store with the default android debug key to sign the Android Application Package (.apk).
The source code, resources and libraries are extracted on the SD card accessible to the user who can then modify the app manually or only use the GUI of the app that allows the icon, name, package, colors and theme to be modified.

Any content added to the assets folder on the SD card can be embedded in the app just like the Java key store.
Combined with the wireless transfer capabilities the app becomes effectively a morphing container that can mutate and virally spread without the need for Internet, thus overcoming Internet kill switches and disruptions due to natural disasters or man made interference.

\section{Conclusions}

We presented a wide range of Android-based social media tools for disaster
communication, privacy, and overcoming Internet kill switches.

No clear winner has emerged among the existing robust social media tools.
For numerous years various groups have worked on differing approaches.
No project has achieved the maturity and popularity of the mainstream fragile
social media tools based on a single profit-driven entity and centralized
architecture. 
Will any robust tool ever reach the popularity of market leaders such as Twitter (300 million
monthly active users)?

Our survey indicates that it is \emph{unlikely} that robustness and resilience will become mainstream.
Existing projects do not seem to be on a sustainable trajectory for mainstream uptake.
The sad conclusions is that the underlying reasons for a stark future are difficult to overcome.
For instance, lack of funding, lack of dedicated expert developers, highly fragmented development effort,
replication of effort instead of software re-use, lack of usability testing, and minimal user experience expertise. 

\IEEEtriggeratref{64}

\bibliographystyle{IEEEtran}
\bibliography{IEEEabrv,LiteratureSurvey}

\end{document}